\documentclass[12pt]{iopart}
\usepackage{bm}

\def\lessim{\lower.5ex\hbox{$\; \buildrel < \over \sim \;$}}
\def\gtrsim{\lower.5ex\hbox{$\; \buildrel > \over \sim \;$}}
\begin{document} \hbadness=10000
%\topmargin -0.8cm\oddsidemargin = -0.7cm\evensidemargin = -0.7cm
%\documentclass[twocolumn,showpacs,amssymb,aps]{revtex4}
%\documentstyle{article}
%\begin{document}
\newcommand{\ea}[1]{\begin{eqnarray} #1 \end{eqnarray}}
\title{Quantum Collective QCD String Dynamics}
\author{Steven Steinke  and  Johann Rafelski}
\address{Department of Physics, University of Arizona, Tucson, Arizona 85721}
\begin{abstract}
The string breaking model of particle production is extended in order 
to help explain the transverse momentum distribution in elementary collisions.
Inspired by an idea of Bialas', we treat the string using a collective 
coordinate approach.
This leads to a chromo-electric field strength which fluctuates, and in turn 
implies that quarks are produced according to a thermal distribution. 
\end{abstract}
%\maketitle
\section{Introduction}
We study the production process of hadronic particles in elementary 
collisions based on the string breaking mechanism of QCD.
The non-Abelian SU(3)-color gauge group leads to confinement of the 
color electromagnetic field lines at scales of roughly 1 fm and above.
Thus, a $q\bar q$ pair produced in, say, an $e^+e^-$ collision will 
have a chromo-electric flux tube connecting the quark and antiquark.
The tube increases in length as the quarks separate, and eventually 
there is sufficient energy in the field to produce another $q\bar q$ pair.
The rate of production is calculated via Schwinger's formula~\cite{Sc1},
generalized to consider transverse particle momentum~\cite{CNN1}.
This is often referred to as the string breaking mechanism of particle production.

Despite its successes, one troublesome detail of the QCD-string
model has not been fully explained.
The predicted transverse momentum distribution of quarks is 
Gaussian~\cite{CNN1}, while the observed spectrum of final state 
particles is thermal, or very nearly so~\cite{RH1,Be1}.
Na\"{\i}ve averaging over constituent quark transverse momenta 
cannot make an exponential distribution out of a Gaussian one.
Further, it is not natural  to suppose that there is sufficient 
rescattering for thermalization in systems as small as those 
in $e^+e^-$ and $p\bar p$ collisions. Hagedorn thus spoke before QCD 
was developed of `pre-established thermal equilibrium'~\cite{RH1}.
While the string model itself has been refined several times, nothing 
like a thermal $p_\perp$ spectrum has been predicted by any of these refinements
until recently.

This discrepancy invites new ideas.  
Bialas showed that a random fluctuation of the string tension could produce 
particles in a thermal spectrum~\cite{Bi1}, if the appropriate initial distribution is used.
He proposed that these fluctuations originate from the stochastic nature of the QCD vacuum.
There has been further work on an extensions of this idea using dynamical 
fluctuations of the string in time~\cite{Flo1}.
This method seems like a promising way to proceed, and the purpose of this 
paper is to improve the justification for these fluctuations, 
as well as to propose a possible  origin.

We separate the transverse and longitudinal dynamics of the string,  
introduce collective coordinates for the transverse dynamics of the string  
and quantize them.
The reason that we will proceed this way is the following order of magnitude 
consideration: due to the very high momentum in the longitudinal direction 
(constituent quarks may have $p_L \approx 100 GeV$ or more, 
in a 1 TeV $p\bar p$ collision) it is reasonable to treat this 
part of the dynamics classically. However, the typical transverse 
momentum of a produced $q \bar q$ pair will be on the 
order of the temperature (which we are seeking to justify) $T\simeq 160$\,MeV.
The radius of the string is about 1\,fm, approximately the same magnitude
as the DeBroglie wavelength of the produced particles.

Our approach is inspired by prior collective coordinate approaches 
in many-body quantum systems.
One well-known example is the excitation spectrum of vibrations and rotations 
of large nuclei\cite{Gr1}.
In the simplest liquid drop version of this model, Coulomb repulsion competes 
with nuclear surface tension.
Quantization of the Hamiltonian resulting from these interactions leads to 
an excitation spectrum which can be verified experimentally.

In the string model, we once again have two competing energy considerations.
The field lines will tend to spread in order to minimize their energy content.
However, the vacuum confines color; this effect can be effectively reproduced 
by implementing in transverse direction the bag pressure and energy density dynamics.
One might ask where the many-body quality arises in the string picture, 
since there is just a single $q \bar q$ pair anchoring the ends of the string.
The answer is in the large number of virtual gluons which constitute the string
QCD fields. We consider the string to be a collective excitation of this gluon sea.

In section \ref{wave}, we first look at the classical string picture,
 namely the energy balance and the Hamiltonian that results.
Quantizing this Hamiltonian yields a harmonic oscillator-type 
wave equation for the transverse dimension of the string.
We explore some of the properties of the solutions to this equation.
Then, in section \ref{spectrum}, we fold the resulting probability 
distribution for the string tension 
with the Schwinger formula for pair production~\cite{Sc1,CNN1}.
This in fact generates an exponential $p_\perp$ spectrum.

%%%%%%%%%%%%%%%%%%%%%%%%%%%%%%%%%%%%%%%%%%%%%%%%%%%%%%%%%%%%%%%%%
\section{\label{wave}Wave Equation}
\subsection{String Hamiltonian}
In preparation for quantization, we will first write down a Hamiltonian, focusing here on 
the energy density per unit length. 
We combine the contributions to the energy density from 
the chromo-electric field and the vacuum pressure to obtain the classical Hamiltonian density
\ea{\label{ham}&\sigma = \frac{1}{2}E_L^2A + BA.}
The two terms are reminiscent of the usual kinetic  and potential energies.
Since the chromo-electric field is longitudinal, we will denote 
it by $E_L$ to avoid confusion with energy.

Suppose  the field lines are broken by a $q\bar q$ pair produced, each having color charge $g$.
In the case where the field is longitudinal, we may simplify the non-Abelian Gauss' law down 
to its usual form:
\ea{\label{gauss}&\int\sum_a\vec E_a\lambda_a \cdot d\vec A = \frac{1}{2}g.}
The factor $\frac{1}{2}$ arises comparing the elementary currents in  QCD and QED: 
\ea{j^\mu_a \equiv \frac{g}{2}\bar\Psi\gamma^\mu\lambda_a\Psi,\qquad
j^\mu \equiv e\bar\psi\gamma^\mu\psi.} 

Note that we have taken $E_L$ to be constant here.
This is easily justified: under the constraint of a 
fixed flux, the total energy is minimized by a uniform field.
Due to the presence of the constraint equation (\ref{gauss}), 
one eliminates the dependence on, say, $E_L$, and obtains:
\ea{&\sigma = \frac{g^2}{8A}+BA.}
We can see the interplay between field and vacuum energy more clearly now.
Minimizing $\sigma$ with respect to $A$ gives
\ea{&A^{\rm classical} = \frac{g}{2\sqrt{2B}}, \\[0.3cm]
&E_L^{\rm classical} = \sqrt{2B},}
and thus a classical string tension of:
\ea{&\sigma = \frac{g}{2}\sqrt{2B} = \frac{g}{2}E_L^{\rm classical}.}

%%%%%%%%%%%%%%%%%%%%%%%%%%%%%%%%%%%%%%%%%%%%%%%%%%%
%\subsection{Surface-Field Collective Dynamics}
%\indent As a first attempt, we simply try $E_L$ and $A$ as canonically conjugate coordinates.
%This is logical since as more energy is stored via the compression of field lines, 
%less is in the vacuum pressure, and vice-versa.
%We will work in the $E_L$ representation, since this is the quantity of most interest to us.
%We assume a commutator of the form
%\ea{[A,E_L] = g}
%Much like the commutator betwen $x$ and $p$ relates to the quantization of action, 
%this equation resembles a quantization condition for color charges.
%Unfortunately, if $E_L$ is Hermitian, $A$ will not be.
%However, adding an $i$ to to the commutator leads to a nonsensical solution.
%Let us overlook this difficulty for the moment.
%Writing down the Schroedinger-like equation gives
%\ea{\sigma\phi = (\frac{1}{4}gE_L+gB\frac{\partial}{\partial E_L})\phi}
%which has solution
%\ea{\phi(E_L) = e^{(8\sigma E_L-gE_L^2)/gB}}
%
%Though a shifted Gaussian looks like a promising distribution, there are two main problems.
%The first is that the Hamiltonian is not Hermitian, for the reasons mentioned.
%The second is that there is a finite probability of negative string tension, 
%which would have to be removed by hand.
%These difficulties lead us to abandon this initial method.
%%%%%%%%%%%%%%%%%%%%%%%%%%%%%%%%%%%%%%%%%%%%%%%%%%%%%%%%%%%%

\subsection{Position-momentum collective dynamics}
The string arises as a quasi particle state (presumably) in the full theory
of QCD. Thus, its dynamics must be treated in an appropriate quantized way.
However, as we noted earlier, the longitudinal evolution of the string should 
be safe to approximate classically, at least in our initial crude treatment.
Furthermore, we assume the string is rigid.
To be precise, there is no spatial bending of the string, nor spatial 
dependence in the field strength.

There are two reasons that we make this assumption.
First, though the energy may fluctuate between bag and field energy, 
we suppose that the configuration of either is such that their separate 
energy contents are minimized.
Second, treating the complexities of a ``wavy'' string is well beyond 
the scope of this paper.
The chromo-electric field now contains only 1 degree of freedom, the longitudinal strength.
Therefore, its conjugate coordinate, which will be the string cross section 
area or something related, will retain only 1 degree of freedom.
Thus, it appears that the initially 2-dimensional transverse dynamics
may in fact be successfully modeled by a 1-dimensional wave equation.
Despite the simplifications, it will still be a problem with a constraint, 
 equation~(\ref{gauss}) (Gauss' law),
and an unusual looking Hamiltonian.

To deal with the transverse dynamics as a quantum problem  in an expedient manner, 
we   make the following further simplification.
As is evident the problem arising should still be closely related to the original one, though, 
admittedly, there is no absolute guarantee of complete isomorphic relation.  
We take the constraint equation~(\ref{gauss}) and substitute it just 
once into the Hamiltonian, equation~(\ref{ham}), yielding:
\ea{\label{qham}&H_\sigma = \frac{1}{4}gE_L+BA.}
We want to quantize this Hamiltonian in terms of Hermitian operators 
with the usual dimensions of distance and momentum.
Since our degrees of freedom are $A$ and $E_L$, 
which dimensions are square of distance and momentum, we introduce
\ea{\label{qop}&x = \sqrt A \\ &p = \sqrt{\frac{g}{2}E_L}.}
Here we have defined $p$ to be the root of the classical energy density.
%This is a bit odd at first, but it saves us from having to worry 
%about radial boundary conditions, 
Since the square root may take 
on positive or negative values, we will allow   $x,p\in(-\infty,+\infty)$.
We make $x$ and $p$ into canonically conjugate operators, 
and so they satisfy the commutation relation
\ea{\label{qopcom}&[x,p] = ig}
The justification for our choice of commutator will follow.
%However, it is known that the harmonic oscillator has a 
%special relation with its classical equivalent.
%Namely, the value of $\hbar$ actually cancels out in the 
%classical limit; everything is expressible in terms of $\omega$.
%Thus, we can at least feel somewhat secure that $C$ will ultimately 
%be of little relevance.
Indeed, $H_\sigma$ now looks like the Hamiltonian of a harmonic oscillator:
\ea{\label{qhamHO}&H_\sigma = \frac{1}{2}p^2+Bx^2.}
Working in the $p$-representation leads to the wave equation:
\ea{\label{HOSch}&\left(\frac{1}{2}p^2 - Bg^2\frac{\partial^2}{\partial p^2}\right)\psi_n
     = \sigma_n\psi_n.}
This is a very familiar equation, and its eigenvalues are:
\ea{\label{HOSchE}&\sigma_n = \sqrt{2Bg^2}\left(n+\frac{1}{2}\right).}
Note that if the dimension of the problem is in fact not 1 but $d$, 
the $\frac{1}{2}$ would have to be multiplied by $d$, and there would be additional $n$'s.

As a matter of convenience, let us define
\ea{\label{Tdef}&T_0 \equiv \sqrt{\frac{\sigma_n}{2\pi(2n+1)}}}
Then we may express the eigensolutions of the problem as
\ea{\label{HOEig}&\psi_n(p) = \frac{A_n}{\sqrt{T_0}}{\mathcal H}_n(\sqrt{p^2/4\pi T_0^2})e^{-p^2/8\pi T_0^2},}
where ${\mathcal H}_n$ are Hermite polynomials, and $A_n$ is a dimensionless normalization.
\subsection{Properties of the quantized string}
Let us now take stock of some of the features that have become evident in our model.
First of all, we may calculate the expected cross sectional area and field strength; because
\ea{\label{equipartition}&\langle\frac{1}{2}p^2\rangle = \langle Bx^2\rangle = \frac{1}{2}\sigma_n}
we obtain
\ea{\label{expE}&\langle E_L\rangle = \frac{2\sigma}{g} = \sqrt{2B}(2n+1),}
\ea{\label{expA}&\langle A\rangle = \frac{\sigma_n}{2B} = \frac{g}{\sqrt{8B}}(2n+1).}
In both cases there is agreement between the ground state and the classical picture.

Now, we want to see if the constraint, as well, is satisfied.
It is reasonable to look at the product of the expectation values and see if Gauss' Law results.
\ea{\label{const1}&\langle E_L \rangle\langle A\rangle = \frac{g}{2}(2n+1)^2}
By using this combination of operators we see that the ground state at least satisfies (\ref{gauss}).
This also clarifies our choice of commutator in equation (\ref{qopcom}).
Whether or not this excludes excited states of the string remains to be seen.
For if we consider different combinations of expectation values that still appear to be 
related to the constraint, we attain different results.
Consider, for example,
\ea{\label{const2}&\langle xp \rangle\langle px\rangle = \frac{g}{2}.}
This should be related to the constraint, and it indicates that all 
excited string states may satisfy equation (\ref{gauss}).
Finding the correct combination of operators to verify if 
the constraint  equation (\ref{gauss}) is satisfied remains 
a difficult problem. However, these preliminary results strongly 
suggest that at least one, if not more, string states are admissible solutions.

%%%%%%%%%%%%%%%%%%%%%%%%%%%%%%%%%%%%%%%%%%%%%%%%%%%%%%%
\section{\label{spectrum}Thermal $p_\perp$ Spectrum}
\indent We are now ready to investigate the spectrum of 
particles produced from this string.
First, let us look at the probability distribution $P(E_L)$.
Taking the probability density $\psi_n^2$ and changing 
coordinates back to the field strength, we obtain
\ea{\label{PEL}& dP_n(E_L) = \frac{B_n}{T_0\sqrt{gE_L}}
    {\mathcal H}^2_n(\sqrt{gE_L/8\pi T_0^2})e^{-gE_L/8\pi T_0^2}dE_L.}
$B_n$ is again a dimensionless normalization.
Note that the factor $(gE_L)^{-1/2}$ arises from the change in measure.
The singularity arising for even $n$ is not troubling; it is an artifact of 
the change in measure and never worse than $E_L^{-1/2}$.

Next, we review the traditional particle production spectrum.
The probability per unit 4-volume of a quark pair of momentum 
$p_\perp$ being produced is given by the Schwinger mechanism, 
as expounded by Casher, Neuberger and Nussinov\cite{CNN1}:
\ea{\label{original}dP(p_\perp) = \frac{gE_L}{8\pi^3}e^{-2\pi m_\perp^2/gE_L}d^2p_\perp.}
This is in fact only the first term in an infinite sum,
but the approximation is appropriate for our current level of precision.
We consider each string breaking to be, essentially, a measurement of $E_L$.
For now, let us suppose that the strings are formed in the ground state, 
with string tension $\sigma_0$.
Thus, over several string breakings, the observed spectrum of produced 
quarks is obtained by folding equations (\ref{PEL}) and (\ref{original}) 
over $E_L$. 
%We can calculate this using the identity
%\ea{\int_0^\infty\frac{1}{\sqrt{x}}e^{-Ax-B/x}dx=\sqrt{\frac{\pi}{A}}e^{-2\sqrt{AB}}.}

Performing this folding results in the quark transverse production probability per unit 4-volume,
\ea{\label{result}dP_{\mathrm {quark}}(p_\perp) = \frac{CT_0^2}{g}(1+m_\perp/T_0)e^{-m_\perp/T_0}d^2p_\perp.}
%The normalization factor $C$ has been omitted for brevity.
This agrees with previous results up to a linear prefactor\cite{Bi1,Flo1}. 
As a final step, consider the composite hadron spectrum.
We will obtain this by first rewriting the distribution in terms of the transverse mass (fortunately, it is already more or less in such a form),
and then folding two distributions together, corresponding to quark-antiquark or quark-diquark pairings.
The final hadron spectrum thus obtained is
\ea{\label{hadronic}\frac{dP_{\mathrm{hadrons}}(m_\perp)}{dm_\perp} & \propto & \left (2X_3-\frac{6}{5}X_5+X_\perp
\left (-3X_2+2X_3+3X_4\right )\right .\nonumber \\
&&\left . -X_\perp^2\left (3X_2+2X_3\right )+X_\perp^3+
X_\perp^4+\frac{1}{5}X_\perp^5 \right )e^{-X_\perp}}
where
\ea{\label{bigm}X_n \equiv \left (\frac{m_1}{T_0}\right )^n+\left (\frac{m_2}{T_0}\right )^n, \qquad X_\perp \equiv \frac{m_\perp}{T_0} \nonumber}
and $m_1$ and $m_2$ are the constituent quark/diquark/antiquark masses.
Note that the largest root of the prefactor is $m_\perp = m_1 + m_2$, and thus the probability is always positive, as expected.
This can be seen trivially in the case $m_1=m_2=0$.
The specific relationship (\ref{Tdef}) is the same as in the aforementioned works, though the pre-factor shifts slightly
the fitted temperature.
Making use of a fairly standard value for the string constant, 
$\sigma_0 \approx 0.9$ GeV/fm, gives $T_0 \approx 165$\, MeV.

\section{Conclusions}
We have considered the string between a $q\bar q$ pair as a 
quasiparticle excitation of the gluon field.
Thus, we justified introducing collective quantization of transverse dynamics.
This implies that the internal chromo-electric field is not constant, but rather when the string
breaks and the field is observed, it takes on 
values according to a quantum probability distribution.
Folding this distribution with the traditional Schwinger-Casher-Neuberger-Nussinov result leads to an intrinsically thermal $p_\perp$ spectrum.
The emerging phenomenology is quite satisfactory; e.g. it produces a thermal spectrum.
However, a more complete theory will certainly lead to further insights including refinement of the temperature parameter.
%This in turn predicts that the total yield of strange to light quarks will be different 
%from the classical string picture to the quantized string picture.

\vspace*{.2cm}
\subsubsection*{Acknowledgments}
Work supported by a grant from: the U.S. Department of Energy  DE-FG02-04ER4131.
SKS thanks SQM2006 for local support.

%%%%%%%%%%%%%%%%%%%%%%%%%%%%%%%%%%%%%%%%%
\vspace*{-0.3cm}
%%%%%%%%%%%%%%%%%%%%%%%%%%%%%%%%%%%%%%%%%


\section{References}
\begin{thebibliography}{99}
\providecommand{\bibinfo}[2]{#2}
\bibitem{Sc1}
  J.~S.~Schwinger,
%  ``On gauge invariance and vacuum polarization,''
  Phys.\ Rev.\  {\bf 82}, 664 (1951)
\bibitem{CNN1}
   A. Casher, H. Neuberger, S. Nussinov {\it Phys. Rev. D} {\underline{20}} 179 (1979)

\bibitem{RH1}
%\cite{Hagedorn:1967ua}
%\bibitem{Hagedorn:1967ua}
  R.~Hagedorn and J.~Ranft,
  % ``Statistical thermodynamics of strong interactions at high-energies. 2.
  %Momentum spectra of particles produced in p p collisions,''
  Nuovo Cim.\ Suppl.\  {\bf 6}, 169 (1968);\\
  %%CITATION = NUCUA,6,169;
%\cite{Hagedorn:1965st}
%\bibitem{Hagedorn:1965st}
  R.~Hagedorn,
  %``Statistical thermodynamics of strong interactions at high-energies,''
  Nuovo Cim.\ Suppl.\  {\bf 3}, 147 (1965);\\
  %%CITATION = NUCUA,3,147;%%
H. Grote, R. Hagedorn and J. Ranft, ``Atlas of Particle Production Spectra'', 
CERN-Black Book Report 1970.
\bibitem{Be1}
  F. Becattini [hep-ph/9701275] (1997),
``Universality of thermal hadron production in $pp$, $p{\bar p}$ and $e^+e^-$ collisions''
in`Proceedings of the XXXIII Eloisatron Workshop "Universality features in Multihadron production and the leading effect", October 19-25 1996, Erice (Italy) 
Report-no: DFF 263/12/1996.

\bibitem{Bi1}
   A. Bialas {\it Phys. Let. B} {\underline{466}} 301 (1999)
\bibitem{Flo1}
   W. Florkowski {\it Acta Phys. Pol. B} {\underline{35}} 799 (2004), and presentation to be published in proceedings of 46th Cracow School of Theoretical Physics (2006)
\bibitem{Gr1}
   see e.g. A.E.S. Green in {\it Nuclear Physics}, McGraw-Hill, New York (1955)
%\bibitem{BeHe1}
%   F. Becattini, U. Heinz {\it Z. Phys. C} {\underline{76}} 269 (1997)
%\bibitem{SDR1}
%   M.D. Scadron, R. Delbourgo, G. Rupp {\it J. Phys. G} {\underline{32}} 735 (2006)
%\bibitem{Cabo}
%   A. Cabo Montes de Oca, M. Rigol Madrazo {\it Eur. Phys. J. C} {\underline{23}} 289 (2002)
%\bibitem{Kne}
%   F. Knechtli {\it Acta Phys. Pol. B} {\underline{36}} 3377 (2005)
%\bibitem{Collab}
%   HPQCD collaboration: MILC collaboration: UKQCD collaboration: C. Aubin, C. Bernard, C. Davies, C. DeTar,
%   Steven Gottlieb, A. Gray, E. Gregory, J. Hein, U. Heller, J. Hetrick, G. Lepage, Q. Mason, 
%   J. Osborn, J. Shigemitsu, R. Sugar, D. Toussaint, H. Trottier, M. Wingate
%   {\it Phys Rev. D} {\underline{70}} 031504 (2004)
%\bibitem{Che}
%   K.G. Chetyrkin, A. Khodjamirian [hep-ph/0512295] (2005)
%\bibitem{Nar}
%   S. Narison [hep-ph/0510108] (2005)
\end{thebibliography}
\end{document}